# Investigation on the Impact of Heat Waves on Distribution System Failures

Andrea Mazza
*Dipartimento Energia "Galileo Ferraris"*
*Politecnico di Torino*
Torino, Italy
ORCID: 0000-0002-0454-9370

Gianfranco Chicco
*Dipartimento Energia "Galileo Ferraris"*
*Politecnico di Torino*
Torino, Italy
ORCID: 0000-0001-7885-8013

Carmen L. T. Borges
*Electrical Engineering Dept.*
*Federal University of Rio de Janeiro*
Rio de Janeiro, Brazil
ORCID: 0000-0001-7935-9566

*Abstract*—This paper discusses some aspects referring to the characterization and modelling of the resilience of distribution systems in the presence of heat waves. The aim is to identify the specific features that can lead to more detailed modelling of the impact of heat waves on the failures that happen in distribution systems. In particular, with heat waves there are differences between the cumulative distribution function of the time to failure in practical cases and in the theoretical reference used for reliability analysis. These differences may be considered to refine the resilience models due to heat waves. Examples taken from real cases are illustrated and commented.

## I. INTRODUCTION

The last two decades experienced an increase of the extreme weather event (EWE) occurrences. In particular, heat waves appear as relatively new phenomena hitting world zones whose climate was usually considered being temperate, such as Europe: the heat wave in 2003 had multiple consequences [1], from the estimated death exceeding 30,000 people (of which 4,000 in Italy) to the decrease of the quantity and quality of harvests (23 million tons of cereal production decrease compared with 2002), soil erosion, forest reduction, and the impact on energy production and infrastructure operation (for example the increase of the grid faults and the reduction of the power production).

Summer 2022 was characterised by three heat waves, in the period June-August: the highest temperature was recorded in Portugal (47°C, at Pinhão) and those phenomena caused a death excess of at least 16 thousands people [2]. Heat waves like the ones reported have been usually classified as High-Impact/Low-Probability (HILP) events: they strongly affect the proper operation of the infrastructures, but their occurrence rate was so low in the past that they were not even considered as sources of fault. However, the reiteration of these phenomena during the year requires an adaptation strategy to be extensively applied, not only to preserve the public health. In fact, the heat waves may affect the proper operation of the energy infrastructure, particularly the electrical grid because: i) the heat waves are extended and widespread in the territory, ii) most of the infrastructure components are aged, and iii) the network design was not based on events of such magnitude, because of their (past) low occurrence frequency.

The longest is the time for which the customers suffer of lack of energy, the highest are the penalties that the Regulatory Authority applies to the Distribution System Operator (DSO), with the exclusion of catastrophic events. Hence, the DSO must guarantee both the reliable and the resilient operation of the grid, by including approaches in the planning able to

This study was carried out within the project EXTRASTRONG (resilience evaluation by EXperimental and TheoRetical ApproacheS in elecTRical distributiON systems with underGround cables) funded by Ministero dell'Università e della Ricerca within the PRIN 2022 program (D.D.104 - 02/02/2022). This manuscript reflects only the authors' views and opinions and the Ministry cannot be considered responsible for them.

consider both normal conditions and HILP events [3]. This aspect results important in the presence of massive investments on the existing grid. For example, in Europe, the share of investments in distribution networks with respect to the total investments on power grids is supposed to continuously grow, from 66% in 2020 to 80% by 2050 [4]. An important part of those investments will refer to urban distribution systems, due to the increasing share of people living in urban areas (approaching 68% of the total population estimated in 2050 [5]).

ENTSO-E (the association of the European Transmission System Operators) defined resilience as the "ability of the system to withstand increasingly extreme system conditions (exceptional contingencies)" [6]. Other institutes/organizations provided their own resilience definitions, such as the Organization for Security and Co-operation in Europe (OSCE) [7], the World Energy Council (WEC) [8], and the Electric Power Research Institute (EPRI) [9]. Moreover, two working groups belonging to CIRED (Congrès International des Réseaux Electriques de Distribution) and CIGRE (Conseil international des grands réseaux électriques) worked for specific definitions with reference to distribution systems and power systems, respectively:

- CIRED working group [10]: "For electrical distribution systems, resilience can be widely defined with respect to system's ability to withstand rare and extreme events (snow storms, hurricanes, earthquakes, terroristic attacks) and quickly recover to its pre-event resilient state." Furthermore, "Resilience for distribution grids is the ability to prepare for and adapt to changing conditions and withstand and recover rapidly from disruptions".
- CIGRE working group [11]: "Power system resilience is the ability to limit the extent, severity, and duration of system degradation following an extreme event." Furthermore, "Power system resilience is achieved through a set of key actionable measures to be taken before, during, and after extreme events such as: 1) anticipation, 2) preparation, 3) absorption 4) sustainment of critical system operations, 5) rapid recovery and 6) adaptation including the application of lessons learnt".

In [12] the authors discussed more in detail the CIGRE definition of resilience and compared it with the definitions of reliability, adequacy and security, by highlighting as important aspects the time evolution, the contingency selection, the relevance of the extreme events, the impact on the system, the acceptability criterion and the modeling techniques.

References [13], [14] and [15] present surveys of techniques and methods for quantifying power system resilience. There is a huge variety of approaches depending on the period of resilience curve under study, between pre-, during or post-event, and also on the objective of the study (anticipation, assessment or enhancement).

This paper focuses on methodologies for assessing power system resilience particularly due to heat wave events. Section II recalls some concepts of reliability and resilience and heat wave modelling. Section III addresses specific aspects of the heat wave phenomena that may be useful to formulate advanced resilience models for heat wave phenomena. Section IV presents some examples taken from real cases. The last section contains the conclusions.

## II. BACKGROUND CONCEPTS AND TECHNIQUES

### A. Reliability and Resilience Evaluation Models

In Italy, with the heat waves in urban areas, increasingly high numbers of faults have been recorded in the distribution network in a relatively limited period [16]. These dense occurrences of faults bring a severe threat to the secure operation of the urban distribution system. Due to the weakly-meshed distribution system structure operated in radial configuration, the single fault in a line can be isolated without losing any customer for a long time. However, the heat wave could cause successive faults happening on close lines in a relatively short time after the first one, i.e., before the first fault is repaired. This could lead to severe blackouts for the customers between the two fault locations.

In the realm of power systems, reliability and resilience are two critical concepts. They are distinguished by the type of events they safeguard against. Reliability primarily focuses on high-probability, low-impact (HPLI) events, which have been the traditional basis for designing and operating power systems. On the other hand, resilience primarily concerns HILP events, which are rare and have the potential to affect a large number of people over extended periods. Investment decisions driven by resilience can significantly differ from those based on traditional security standards like *N-1* or *N-2* and reliability-driven investments [13].

Reliability engineering typically focuses on evaluating low to moderately impactful, high-frequency events that may be numerous but affect a limited number of people for short durations, often measured in minutes to hours. In contrast, resilience considerations deal with rare, high-impact events that unfold over extended timeframes. Recent decades have seen a notable increase in the frequency and severity of extreme weather events, such as hurricanes, ice storms, floods, and droughts, primarily due to global climate change. Consequently, electric power systems, particularly those reliant on renewable sources, are increasingly vulnerable to more severe weather-related impacts.

The IEEE PES task force [13] established a definition of resilience that both acknowledges and differentiates itself from the definitions put forth by organizations such as CIGRE, FERC, and various government and industry bodies:

"Power system resilience is the ability to limit the extent, system impact, and duration of degradation in order to sustain critical services following an extraordinary event. Key enablers for a resilient response include the capacity to anticipate, absorb, rapidly recover from, adapt to, and learn from such an event. Extraordinary events for the power system may be caused by natural threats, accidents, equipment failures, and deliberate physical or cyber-attacks."

*B. Modelling Heat Waves and the Related Effects*

Modelling the occurrence and effects of the heat waves on the operation of distribution system components (in particular cables and joints) is a challenging task. The synergistic application of notions coming from different conceptual fields and backgrounds is required, from large-scale climatic models to detailed models of cables and joints that also consider the type of installation.

The literature and the technical developments in the resilience field still have to provide a consistent framework for characterizing the resilience due to heat waves, with models validated in a sufficient number of applications worldwide.

Some concepts classically adopted in the past are now challenged by the evolving consequences of climate change, with the occurrence of extreme events also in areas in which the situation was usually considered less exposed to resilience issues. Examples are the rules of thumb used in different jurisdictions to characterize abnormal events, in which a periodicity of a given number of years (e.g., seven years) was adopted to represent possible time sequences in which extreme temperatures are reached. While this periodicity was somehow accepted to represent past situations, more recently there have been some years characterized by extreme events with periodicity well below to seven years, even with extreme temperatures and heat waves occurring in successive years. In the current context, assuming any periodicity can be abandoned, to follow new types of assessments based on a mix of model-based and data-driven considerations.

In practice, an integrated approach is needed to establish the links between the occurrence of the heat waves and the effects of the heat waves on the distribution networks.

The first source of information comes from climatic models, which aim at providing spatial information at different scales, where the integration of regional climate models adds more local details depending for example on orography [17]. Based on the integrated information, indicators such as the dimensionless Heat Wave Magnitude Index-daily (HWMId) [18] that consider heat wave duration and intensity have been used in various studies to represent the occurrence of heat waves. The HWMId indicator has also been used in simulations carried out to compare the results of different models that combine global and regional climate models, with both re-analysis of historical heat waves and projections for future changes in heat wave magnitudes [17]. The definition of the heat wave period presented in [18] is based on a daily threshold computed as the $^{th}$ percentile of the maximum daily temperature recorded in the reference period 1981-2010 considering the day under analysis centered on a 31-day window. Based on this threshold, the heat wave is conventionally defined if there are three or more consecutive days in which the maximum daily temperature exceeds the daily threshold. Seasonal effects may be considered by assuming that the heat waves could occur only in a restricted part of the year, e.g., from May to September [19]. More inputs may come from considering the effects of rainfalls associated with temperatures [20].

The second aspect is the formulation of dedicated models to assess of the response of the ground to heat wave effects. This requires the analysis of the composition of the soil and of the mechanisms that drive the temperature change in the ground where the cables are installed [21]. Further information comes from the study of the evolution of the electrical demand during time [22], in particular, considering the correlations between the occurrence of heat waves, the increase of the electrical demand during day and night [23] and the specific dependence on temperature of the additional electrical demand activated at very high temperatures. In addition, different types of intermittence or periodicity of the demand, due to seasonal, weekly and daily effects that appear in the demand patterns, have to be considered in conjunction with the thermal dynamics of the ground located around the cable.

For the distribution network, the temporal evolution of the temperature in the cables and joints depends on the combination of the external effects (depending on the characteristics and response of the soil) to temperature variations (depending on the ground characteristics and type of installation) with the internal effects due to the current that flows in the cable.

The last part of the study is more local and refers to the identification of the mechanisms with which the cables and in particular the joints evolve towards the occurrence of failures.

*C. Resilience Models for Heat Wave Analysis*

Non-sequential Monte Carlo (NSMC) simulations serve as a valuable tool for evaluating the spatial impacts of extreme events on power systems. These simulations have been applied both independently and in conjunction with other methods, such as Markov chains and Kantrovich distance-based scenario reduction, to assess the repercussions of failure events on power system resilience [15].

In a study described in [24], the assessment of power system resilience using NSMC is a two-step process. The first step involves determining the state transition of a power system under extreme events based on Markov chains, and the second step calculates resilience indices, considering changes in the network topology following extreme events.

It is important to note that Monte Carlo simulations could not effectively capture rare and extreme events unless specific techniques, such as importance sampling, are employed.

The different conceptual aspects between reliability and resilience have been established under the concepts of HILP and HPLI. However, in the typical comparisons, resilience is intended as the occurrence of extreme events that happen in a given moment and cause immediate significant damages in a large portion of the system. In this respect, a heat wave has some peculiarities that differ from other extreme events.

For example, in case of a cable or joint the damage in general does not happen immediately after the beginning of the extreme event but occurs with a delay that depends on the evolution of the various phenomena that cause the disruption of the insulating material. In addition, by carrying out accurate studies on the evolution of the climate in a specific area, the probability of occurrence of the events over the entire distribution system could not be considered as extremely low.

The impact itself depends on the possibility of anticipating the possible occurrence of failures in periods when the heat wave is expected to occur. All these aspects could create a stronger link between reliability concepts and the aspects referring to heat wave phenomena with respect to what happens with many other extreme events.

Establishing a threshold between failure events happening on distribution system cables and joints that can be associated to reliability or to resilience (due to heat waves) is a challenging aspect that cannot be solved with a "rigid" definition. The main aspects to consider include the number of events per day, the delay between periods with extreme temperatures and the failure, due to the evolution in time of thermal and mechanical phenomena not easy to be modelled. Hence, failure events depend on the past history of the cable or joint operation, an aspect that is outside the conceptual reliability framework.

Some "rigid" definitions included in the current standards or in the regulation can be easily questioned (8 hours, minimum number of failure events to be associated with resilience). For these reasons, conceptually the techniques of analysis used for reliability calculations cannot be directly adapted to study the failures depending on resilience issues due to heat waves.

III. TOWARDS A REFINED RESILIENCE MODEL FOR REPRESENTING HEAT WAVE PHENOMENA

Starting from the general scheme for analyzing resilience, specific models are needed to assess the resilience after heat waves. In fact, the nature of the heat wave-related phenomena introduces dedicated aspects to be considered, such as:
- *Fatigue*: the response of cables and joints is not instantaneous, it requires time to follow the evolution of the thermal transients that lead to failures. This aspect can be related to the fatigue due to the evolution of the thermal quantities, related to the properties of the materials.
- *Delay*: because of fatigue aspects, the failures in cables and joints occur with some delay with respect to the cause.

Moreover, the failures do not occur all together but are delayed in a substantially poorly predictable way because of the intrinsic characteristics of the materials and of the propagation of the internal defects that lead to failure. As such, the delay is not totally random. More refined models of the cable thermal behavior are needed to incorporate the effects of failure occurrence and propagation also depending on fatigue and the time evolution of the cable loading.

- *Successive failures*: resilience issues due to fatigue occur when many cables fail in a close period after the beginning of the occurrence of heat wave conditions. With respect to the classical reliability studies, in which the time between failures (TBF) is a relevant quantity and the probability distribution of the TBF for single independent faults conceptually follows the exponential distribution, the presence of heat wave conditions introduces different aspects. First of all, the failures due to a heat wave could be considered as failures with common cause. However, this common cause is different with respect to the common cause traditionally taken into account in reliability studies, i.e., causes that appear at the same time due to a sudden common event (e.g., weather-related, or failures in substation equipment that cause dependent outages) or involve side effects of the action of the protection systems or human errors [25], with failures that appear within a relatively close time period (from seconds to minutes). For heat waves, the timing of the failures is generally longer (up to hours) and depends on the past history of operation of the cables and joints. Moreover, the TBF due to heat waves could be practically shorter than the TBF for single independent failures considered in the reliability study.

IV. APPLICATION EXAMPLES

This section presents an analysis that takes as inputs the faults occurred on the distribution system in a city in the North-West of Italy in the last years. Fig. 1 shows the Empirical Cumulative Distribution Function (ECDF) of the TBF for the pairs of years 2001-2002 and 2017-2018, together with the exponential Cumulative Distribution Function (CDF) with the parameter mean time between failures (MTBF) of the faults occurred over the entire distribution system.

Regarding the faults occurred in the years 2001-2002, with data taken from [26] that refer to a portion of the distribution system in service in these years, the shape of the ECDF and the shape of the exponential CDF are quite similar, indicating that most of the faults can be considered basically independent of each other; hence, the classical reliability approach can be used. Indeed, no specific period with excessively high temperatures appeared in these two years. Conversely, the ECDF for the years 2017-2018 exhibits a worst match with the exponential CDF: in particular, the ECDF highlights a non-negligible number of faults with low TBF, which affect the system more than expected. This phenomenon cannot be totally explained through the classical reliability approach. In 2017-2018, some heat wave phenomena occurred and can be a cause of discrepancy between the ECDF and the exponential CDF at relatively small TBF values.

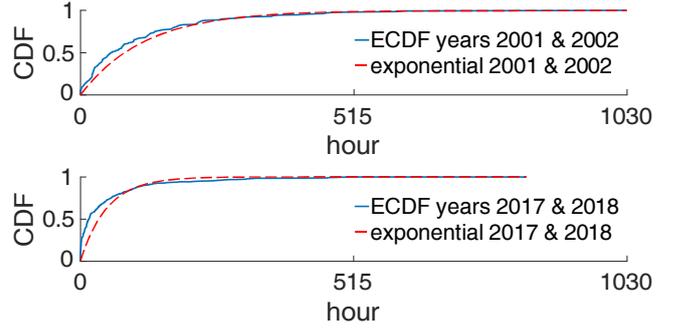

Fig. 1. ECDF and exponential CDF of the TBF related to the years 2001-2002 (on the top) and 2017-2018 (on the bottom).

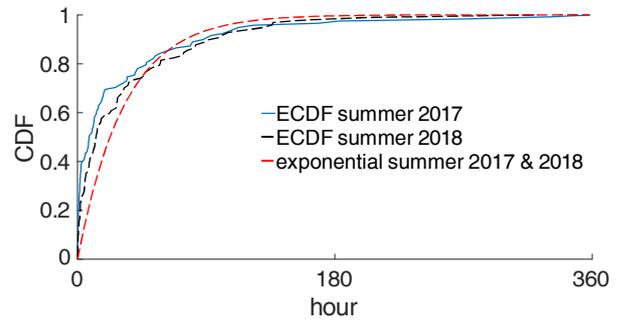

Fig. 2. ECDF and exponential CDF for the period May-September of the years 2017 and 2018.

To clarify this aspect, focusing on the data of the summer period May-September 2017, Fig. 2 compares the ECDF and the exponential CDF of the TBF. There is an evident impact of faults with short TBF, especially in the year 2017. This is aligned with the long-standing heat waves registered during that year (see [16]).

Focusing on the year 2017, Fig. 3 shows the temperatures

(maximum, average and minimum) and the number of daily faults. Two days (#145 - May 25th, and #177 - June 26th) have a particularly high number of daily faults (9 for each day). For these two days, the occurrence of faults during the hours of the day is shown in Fig. 4. Interestingly, not all the faults happen during the day-time (when the temperature reaches the maximum value). This indicates somehow the existence of a delay between the time at which the maximum temperature is reached and the instant of occurrence of the (multiple) faults.

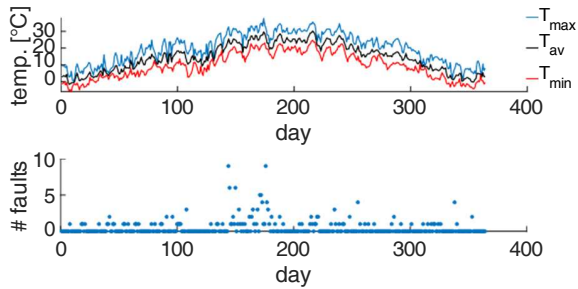

Fig. 3. Temperatures and number of faults (year 2017).

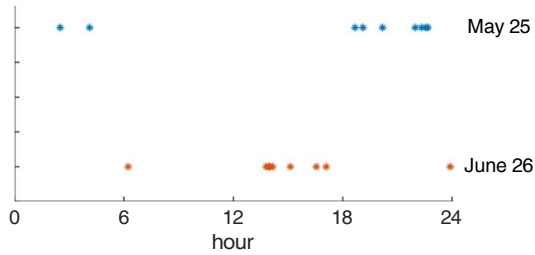

Fig. 4. Fault occurrences on 25 May 2017 and 26 June 2017.

## V. CONCLUSIONS

Specific models are needed to assess the distribution system resilience after heat waves. The identification of failures associated with a heat wave can be conducted by considering together the information on the evolution in time of the ambient temperature, the loading of the electrical lines, an anomalous number of failures in a given period with respect to what happens in normal operating conditions, and the fact that considering the failures due to heat waves as independent failures within a reliability study would lead to a probability distribution of the TBF with an intensification of short periods with respect to the exponential distribution that would be followed without the occurrence of the heat wave. The next steps will include a specific work to identify a "reference temperature" for the area under analysis to identify the heat waves periods (also considering the different definitions existing in the literature). These periods will be correlated with the faults, in order to identify the most relevant features for modeling their impact on the distribution system. Moreover, within the project EXTRASTRONG [27], a specific test-bench will be developed to enable the laboratory replication of the heat waves, in order to provide at DSOs and cable manufacturers specific indications on how verifying the behavior of joints and cables in these particular conditions.